\documentstyle[editedvolume,psfig]{crckapb} 

  at 12.0 true pt 

\def\etal{{et al}}

\def\phoenix{{\tt PHOENIX}}

\def\b{\beta}

\def\rout{\ifmmode{r_{\rm out}}\else\hbox{$r_{\rm out}$}\fi}
\def\tmax{\ifmmode{\tau_{\rm max}}\else\hbox{$\tau_{\rm max}$}\fi}
\def\tstd{\ifmmode{\tau_{\rm std}}\else\hbox{$\tau_{\rm std}$}\fi}
\def\vmax{\ifmmode{v_{\rm max}}\else\hbox{$v_{\rm max}$}\fi}
\def\muE{\ifmmode{\mu_{\rm E}}\else\hbox{$\mu_{\rm E}$}\fi} 
\def\pE{\ifmmode{p_{\rm E}}\else\hbox{$p_{\rm E}$}\fi} 
\def\bmax{\ifmmode{\b_{\rm max}}\else\hbox{$\b_{\rm max}$}\fi}
\def\kms{\hbox{$\,$km$\,$s$^{-1}$}}

\def\ang{\hbox{\AA}}

\def\Teff{\hbox{$\,T_{\rm eff}$} }

\def\alog#1{\times 10^{#1}}

\def\rout{\hbox{$r_{\rm out}$} }

\def\chistd{\ifmmode{\chi_{\rm std}}\else\hbox{$\chi_{\rm std}$}\fi}

\def\lstar{\ifmmode{\Lambda^*}\else\hbox{$\Lambda^*$}\fi} 
\def\Rop{\ifmmode{[R_{ij}]}\else\hbox{$[R_{ij}]$}\fi}

\def\Rji{\ifmmode{[R_{ji}]}\else\hbox{$[R_{ji}]$}\fi}
\def\Rstar{\ifmmode{[R_{ij}^*]}\else\hbox{$[R_{ij}^*]$}\fi}

\def\Rjistar{\ifmmode{[R_{ji}^*]}\else\hbox{$[R_{ji}^*]$}\fi}
\def\DRji{\ifmmode{[\Delta R_{ji}]}\else\hbox{$[\Delta R_{ji}]$}\fi}
\def\DRij{\ifmmode{[\Delta R_{ij}]}\else\hbox{$[\Delta R_{ij}]$}\fi}

\def\ns{\ifmmode{N_{\rm s}}          
        \else\hbox{$N_{\rm s}$}\fi}

\def\mat#1{{\bf #1}}     
\def\vek#1{{#1}}         

\newcount\eqcount
\def
  \nummer{
    \global\advance\eqcount by 1
    (\the\eqcount)
  }

\def
  \numadv{
    \global\advance\eqcount by 1
  }

\def
   \numout#1{
     (\the\eqcount #1)
  }

\def\ivek#1#2{\ifmmode{\vek{I}^{#1}_{#2}}
        \else\hbox{$\vek{I}^{#1}_{#2}$}\fi}

\def\tmat#1#2{\ifmmode{\mat{t}^{#1}_{#2}}
        \else\hbox{$\mat{t}^{#1}_{#2}$}\fi}
\def\rmat#1#2{\ifmmode{\mat{r}^{#1}_{#2}}
        \else\hbox{$\mat{r}^{#1}_{#2}$}\fi}
\def\bvek#1#2{\ifmmode{\beta^{#1}_{#2}}
        \else\hbox{$\beta^{#1}_{#2}$}\fi}

\def\lp{\ifmmode{\lambda^+_\tau}           
        \else\hbox{$\lambda^+_\tau$}\fi}
\def\lm{\ifmmode\lambda^-_\tau             
        \else\hbox{$\lambda^-_\tau$}\fi}

\begin{opening}

\title{NLTE Model Atmospheres for M Dwarfs and Giants\vskip -1cm}

\author{\relax Peter H. Hauschildt}
\institute{Dept.\ of Physics and Astronomy, Arizona State University, Box 
871504, Tempe, AZ 85287-1504\\
E-Mail: \tt yeti@sara.la.asu.edu }

\author{France Allard \& David R. Alexander}
\institute{Dept.\ of Physics,\\ Wichita State University,
Wichita, KS 67260-0032 }

\author{Andreas Schweitzer}
\institute{Landessterwarte Heidelberg, K\"onigstuhl, D-69117 Heidelberg}

\author{E. Baron}
\institute{Dept. of Physics and Astronomy, University of Oklahoma
440 W. Brooks, Rm 131, Norman, OK 73019-0225}

\end{opening}

\runningtitle{NLTE Model Atmospheres for M stars}

\begin{document}
\vskip -0.5cm
\bibliographystyle{apj}






\section{Introduction}

The atmospheres of M stars are dominated by a small number of very strong 
molecular compounds (H$_2$O, TiO, H$_2$, CO, VO). Most of the hydrogen is 
locked in molecular H$_2$, most of the carbon in CO; and H$_2$O, TiO and VO 
opacities define a pseudo-continuum covering the entire flux distribution of 
these stars. The optical ``continuum'' is due to TiO vibrational bands which
are often 
used as temperature indicators for these stars. These may be the depth of 
the bands relative to the troughs in between them; or the depth of the VO 
bands; or of the atomic lines relative to the local ``continuum''; or even the 
strength of the infrared water bands; all of these depend on the strength of 
the TiO bands and the amount of flux-redistribution to longer wavelengths 
exerted by them. Departures from LTE of the Ti~I atom, and thus the 
concentration of the important TiO 
molecule, could, therefore, have severe and measurable consequences on the 
atmospheric structure and spectra of these stars. 

Due to the very low electron temperatures, the electron density is extremely 
low in M stars; even lower than in novae and SNe.  Collisions with particles 
other than electrons, e.g., the H$_2$ or helium, are by far not as effective 
as electron collisions, both because of their smaller cross-sections and of 
their much smaller relative velocities.  Therefore, collisional rates which 
tend to restore LTE, could be very small in cool stars. This in turn could 
significantly increase the importance of NLTE effects in M stars when compared 
to, e.g., solar type stars. 

In this paper we discuss NLTE effects of Ti~I in fully
self-consistent models for a few representative M/Brown dwarf
and M giant model atmospheres and spectra. 

\begin{figure}[t]
\vskip -0 cm
\caption[]{\label{ti1grot}Simplified Grotrian diagram of our Ti~I model atom.  
All 395 levels and 5379 primary (i.e., full NLTE) 
lines are shown but the 0.8 million secondary (approximate) NLTE lines have been 
omitted for clarity.}
\vskip -12pt
\end{figure}

\begin{figure}[t]
\vskip -0 cm
\begin{minipage}[t]{6cm}
\end{minipage}
\begin{minipage}[t]{6cm}
\end{minipage}
\caption[]{Ti~I departure coefficients as function of optical
depth for a NLTE model
with $\Teff=2700\,$K and $\log g=5.0$ (left hand side) and with
with $\Teff=4000\,$K and $\log g=5.0$(right hand side).}
\vskip -12pt
\end{figure}

\begin{figure}[t]
\vskip -0 cm
\begin{center}
\begin{minipage}[t]{10.5cm}
\psfig{file=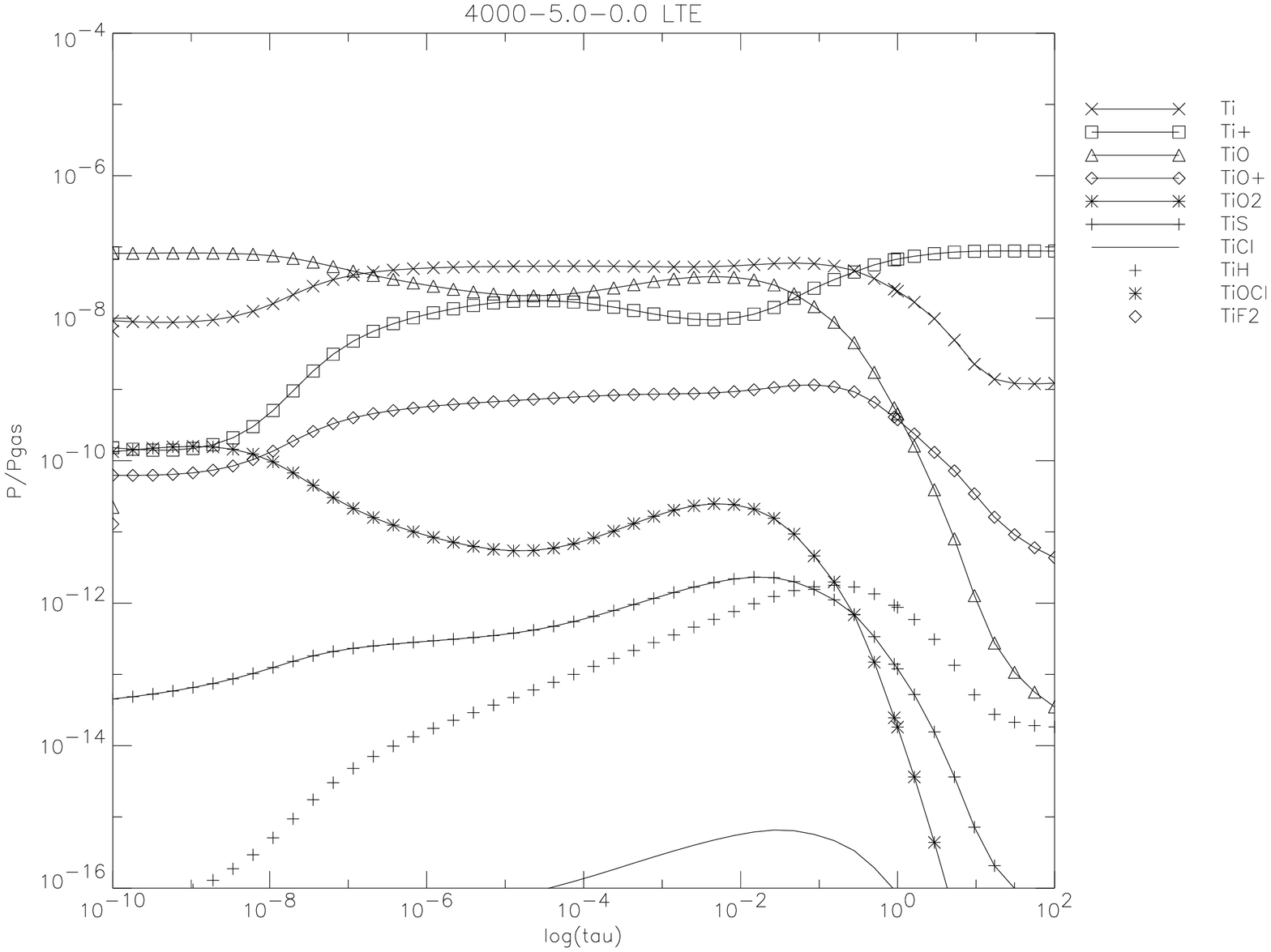,width=10.2cm,angle=90}
\psfig{file=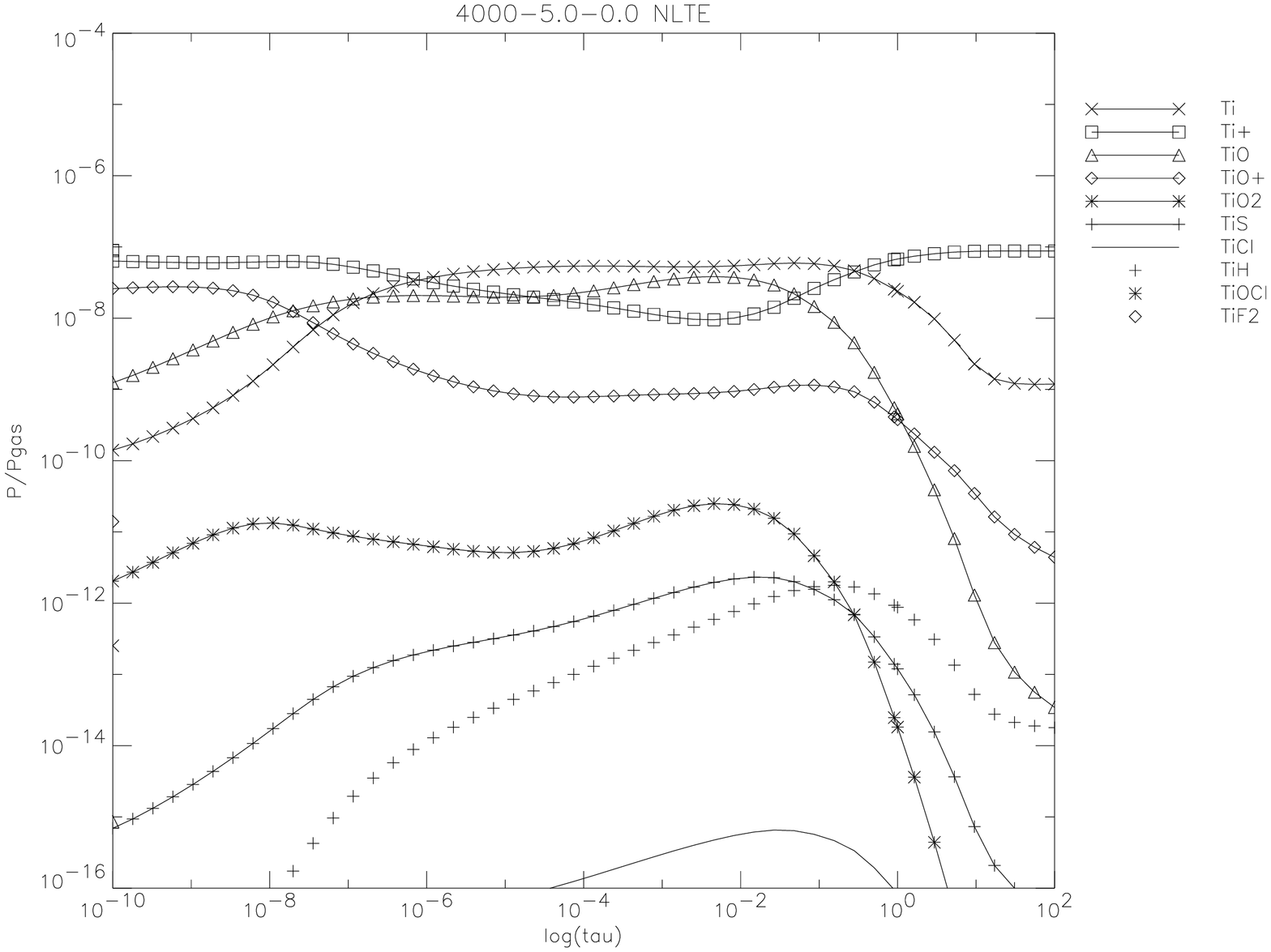,width=10.2cm,angle=90}
\end{minipage}
\end{center}
\caption[]{\label{fig3}NLTE effects on the Ti ionization and molecule formation
for the $\Teff=4000\,$K, $\log g=5.0$ model.}
\vskip -12pt
\end{figure}

\section{Methods and Models}

In order to investigate if Ti~I NLTE effects are important for the formation of
cool star spectra, a full NLTE model calculation is required. This means that
the multi-level NLTE rate equations must be solved for a number of species
self-consistently simultaneously with the radiative transfer and energy equations, including
the effects of line blanketing and of the molecular equation of state. For the
purpose of this analysis we use the model code \phoenix\ originally developed
for the modeling of the expanding atmospheres of novae and supernovae, and
adapted to conditions prevailing in cool stars by Allard and Hauschildt
(1995\nocite{mdpap}, hereafter AH95).
\phoenix\ (version 5.9) uses a spherical radiative transfer
             for giant models (logg<3.5), and an equation of state (EOS)
             recently upgraded from 98 to 195 molecules using the polynomial
             partition functions by Irwin (1998).  For this work, the EOS
             as been extended to include:
\begin{enumerate}\itemsep=0pt\parsep=0pt
\renewcommand{\labelenumi}{(\alph{enumi})}
\item TiO$^+$ and ZrO$^+$ 
         (Gurvich \& Glushko, 1982)\nocite{TiOplus}.
\item H$_3^+$ (and H$_2^+$) to the EOS using the new  H$_3^+$
         partition function by Neale \& Tennyson (1995)\nocite{h3p}. 
\end{enumerate}
 The opacity data base of Phoenix includes assorted f-f processes, 
H$^-$ absorption, Collision Induced Absorption (Borysow 1993), 
JOLA bands for CaH, VO, and FeH (AH95), dust absorption (experimental)
and a direct line-by-line treatment for all available absorbers 
(inculding isotopes):
\begin{enumerate}\itemsep=0pt\parsep=0pt
\renewcommand{\labelenumi}{(\alph{enumi})}
\item atomic lines from updated Kurucz CD\#1
\item J{\o}rgensen TiO \& CN lines, H$_2$O (opacity sampling tables)
\item Goorvitch CO lines
\item the inclusion of the IR roto-vibrational system of SiO using the
         detailed line list by Goorvitch.
\item   H$_3^+$ lines by Neale \& Tennyson (1995)       
\item Kurucz CD\#15 molecular lines
\item Miller \& Tennyson H$_2$O lines for high-res IR spectra
\item Improved water vapor lines by Viti \etal\ (in preparation) as 
soon as they become available.
\end{enumerate}

 For the strongest ca. $6\alog{6}$ atomic \& molecular lines, we use detailed
depth-dependent Voigt profiles with improved damping constant computation
by Schweitzer \etal\ (in preparation), and Gaussian profiles for an 
additional $10\alog{6}$ much weaker lines. 
In addition, we include ca.\ 2000 photo-ionization
cross sections (Verner \& Yakovlev 1995)\nocite{verner95}.
Both the NLTE and LTE lines for atoms and molecules are treated with a direct
line-by-line (LBL) method. This method is
different from the classical opacity sampling (OS) approach in that we do {\em
not} use pre-computed tables for the line opacity as a function of temperature
using only fixed Gaussian line profiles. The LBL method dynamically selects the relevant
lines from the master line lists at the beginning of each iteration and sums
the contribution of every line within a search window to compute the total line
opacity at {\em arbitrary} wavelength points. The latter is important in NLTE
calculations in which the wavelength grid is both irregular and variable (from
iteration to iteration due to changes in the physical conditions). The LBL
approach also allows detailed and depth dependent line profiles to be used
during the iterations. However, in order to make the LBL method
computationally feasible, modern numerical techniques, e.g., block algorithms
with high data locality, and high-end workstations or supercomputers must be
used.
In the calculations we present in this paper, we have set the micro-turbulent
velocity $\xi$ to $2\kms$. We include LTE lines if they are stronger than a
threshold $\Gamma\equiv \chi_l/\kappa_c=10^{-4}$, where $\chi_l$ is the
extinction coefficient of the line at the line center and $\kappa_c$ is the
local b-f absorption coefficient.


\phoenix\ is a full multi-level NLTE code, i.e., the NLTE effects are included
self-consistently in the model calculations. Hauschildt \& Baron
(1995\nocite{fe2pap}) have extended the numerical method
developed by Hauschildt (1993)\nocite{casspap} for NLTE calculations 
with a very detailed model
atom of Fe~II. In this section we describe how we apply this technique to add
the Ti~I atoms to the list of NLTE species already available in the
model calculations.

\begin{figure}[t]
\vskip -0 cm
\begin{minipage}[t]{6cm}
\psfig{file=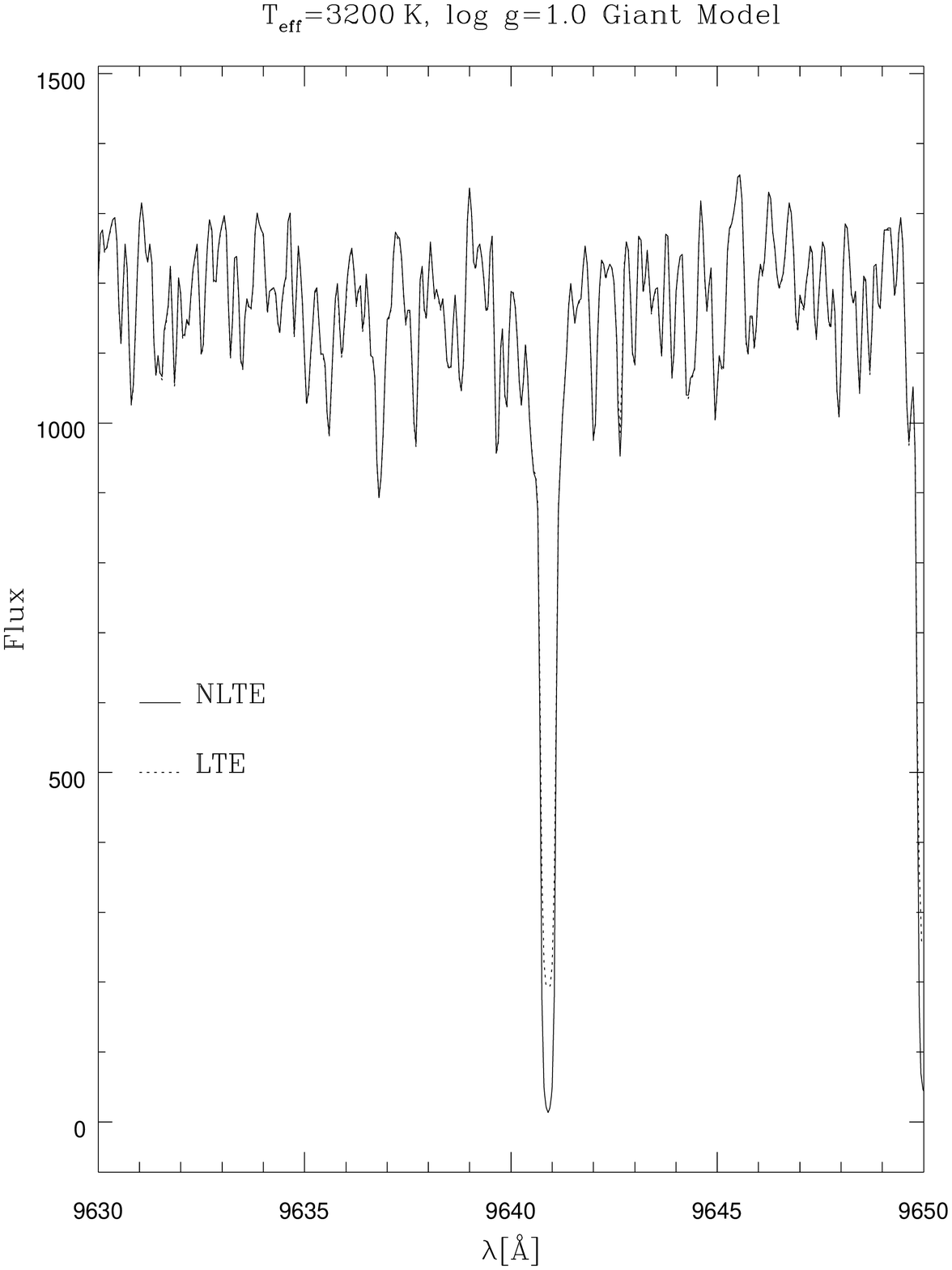,width=6cm,clip=,angle=0}
\end{minipage}
\begin{minipage}[t]{6cm}
\psfig{file=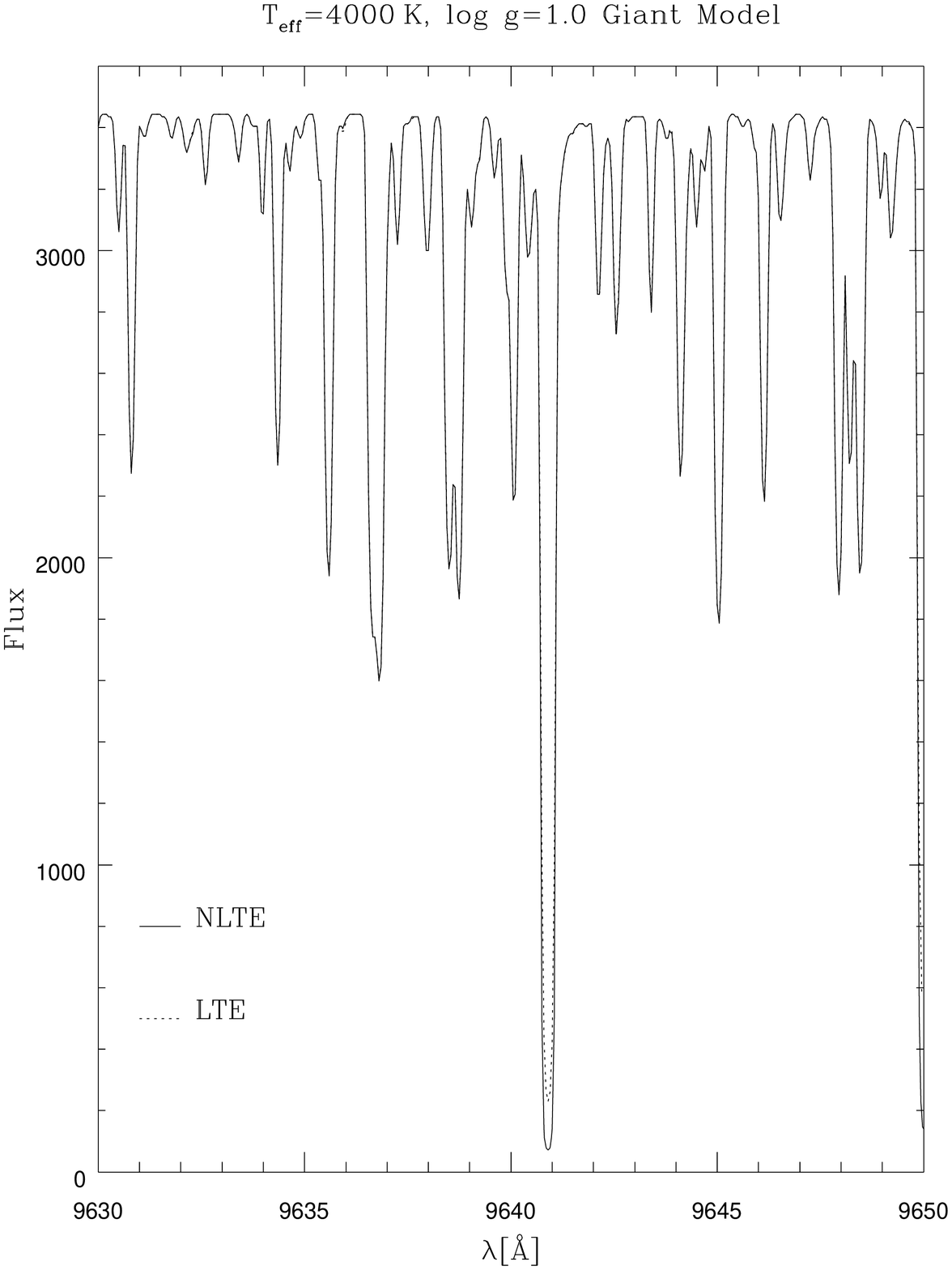,width=6cm,clip=,angle=0}
\end{minipage}
\caption[]{\label{spec3}NLTE effects on the Ti~I line at $\lambda_{\rm vac}
\approx9641\ang$
for two giant models. The model parameters are $\Teff=3200\,$K, 
$\log g=1.0$ for the left hand panel and $\Teff=4000\,$K, $\log g=1.0$ for the right 
hand panel. The LTE spectrum uses the same model structure as the NLTE 
spectrum but with all departure coefficients set to unity.}
\vskip -12pt
\end{figure}

\begin{figure}[t]
\vskip -0 cm
\begin{minipage}[t]{6cm}
\psfig{file=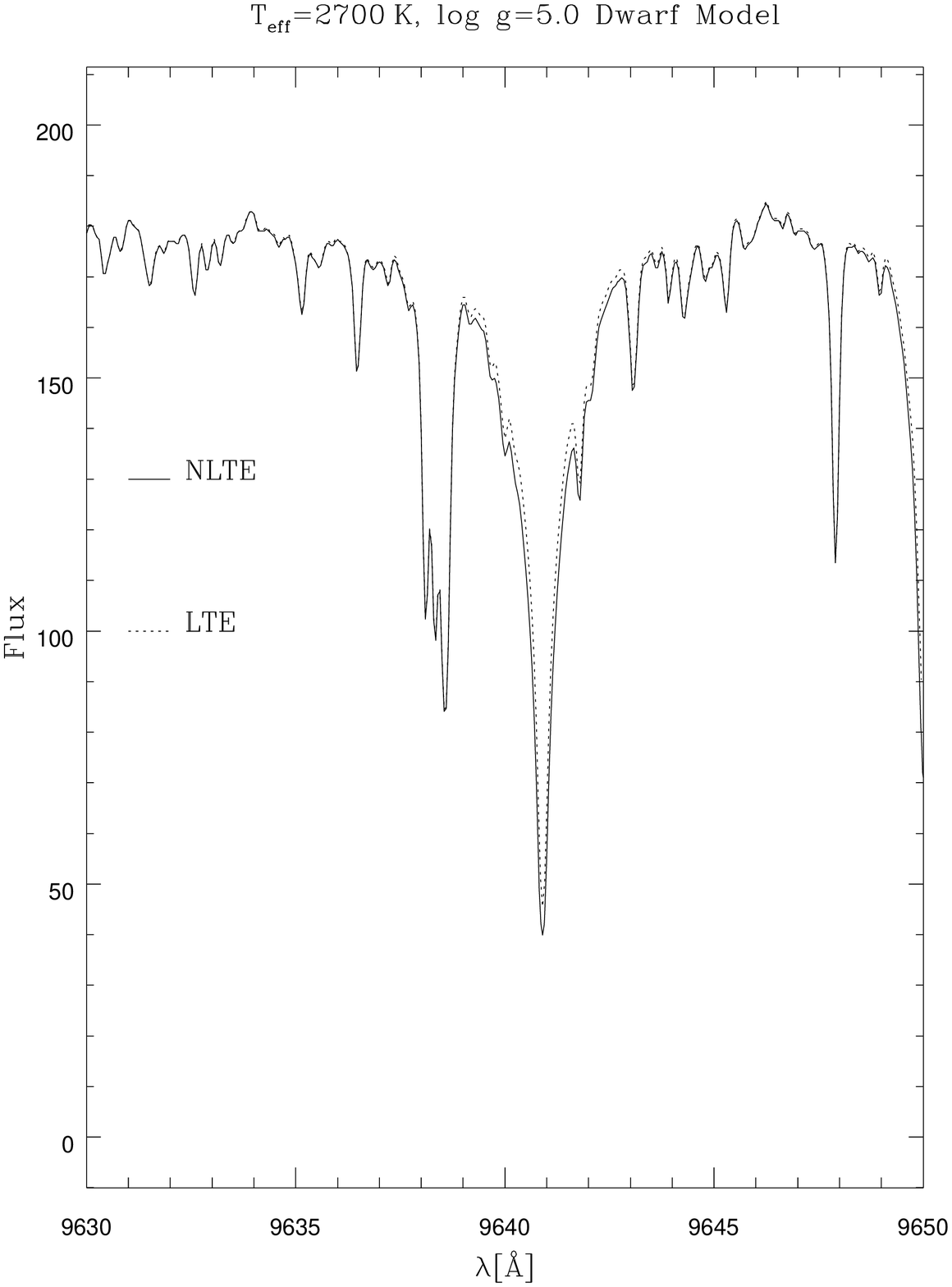,width=6cm,clip=,angle=0}
\end{minipage}
\begin{minipage}[t]{6cm}
\psfig{file=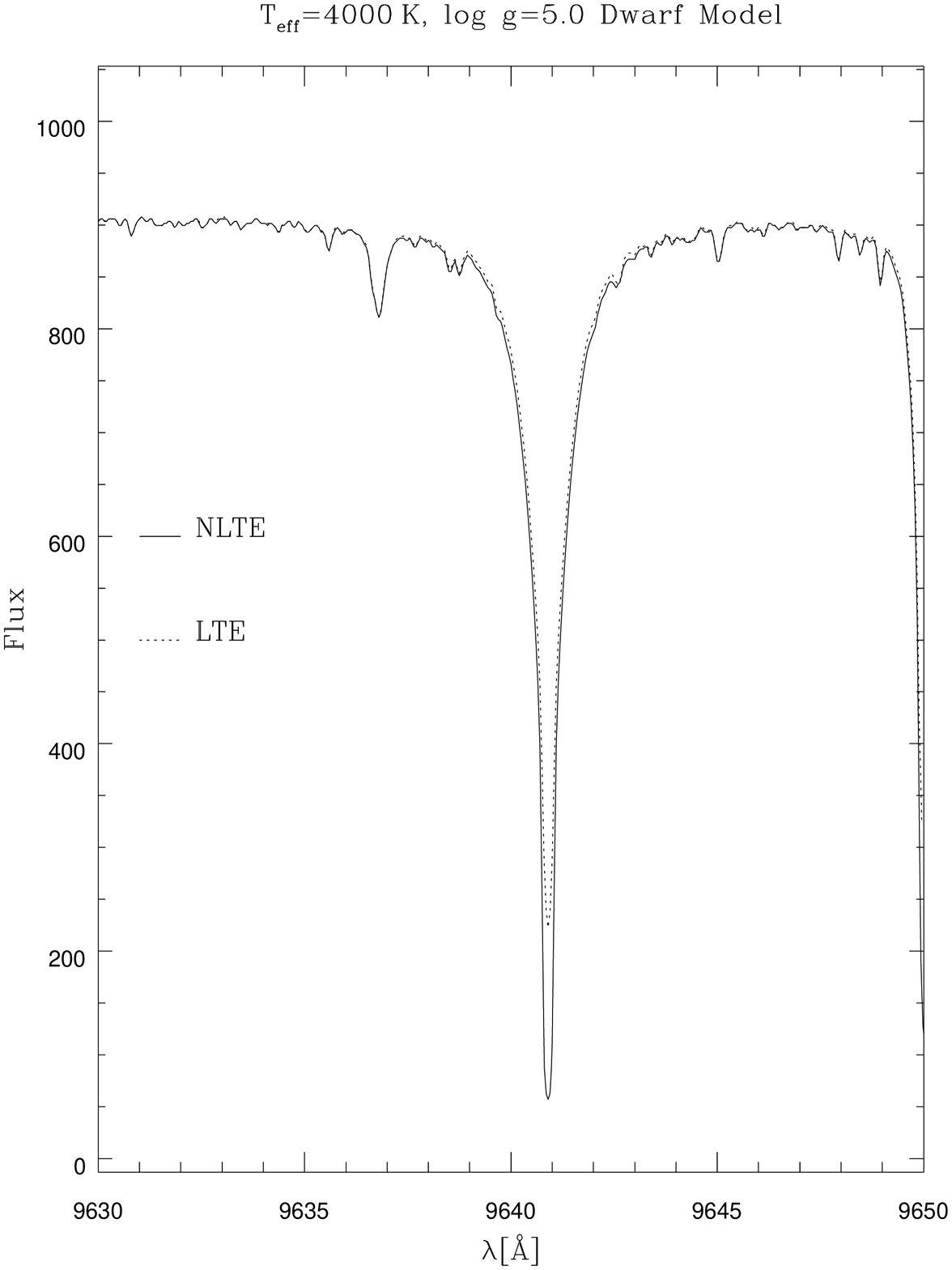,width=6cm,clip=,angle=0}
\end{minipage}
\caption[]{\label{spec4}NLTE effects on the Ti~I line at $\lambda_{\rm vac}
\approx9641\ang$ for
two dwarf models. The model parameters are $\Teff=2700\,$K, 
$\log g=5.0$ for the left hand panel and $\Teff=4000\,$K, $\log g=5.0$ for the right 
hand panel. The LTE spectrum uses the same model structure as the NLTE 
spectrum but with all departure coefficients set to unity.}
\vskip -12pt
\end{figure}

\subsection{The Ti~I model atom}

To construct the Ti~I model atom we have selected the first 34 terms of Ti~I.
We include all observed levels that have observed b-b
transitions with $\log{(gf)} > -3.0$ as NLTE levels where $g$ is the statistical
weight of the lower level and $f$ is the oscillator strength of the
transition. This leads to a model atom with 395 levels and 5279 primary
transitions treated in detailed NLTE. That is, we solve the complete b-f \&
 b-b radiative transfer and rate equations for all these levels including all
radiative rates of the primary lines. A Grotrian diagram of this model atom is
shown in Fig.~\ref{ti1grot}. In addition, we treat the opacity and emissivity
for the remaining nearly $0.8$ million ``secondary'' b-b transitions in NLTE,
if one level of a secondary transition is included in the model. A detailed
description of the method is given in Hauschildt \& Baron (1995).

Photo-ionization and collisional rates for Ti~I are not yet available.
Thus, we have taken the results of the Hartree Slater central field
calculations of Reilman \& Manson (1979)\nocite{reilman79} to scale
the ground state photo-ionization rate and have then used a hydrogenic
approximation for the energy variation of the cross-section. Although
they are only very rough approximations, the exact values of the b-f
cross-sections are not important for the opacities themselves which
are dominated by known b-b transitions of Ti~I and other species.
They do, however, have an influence on the actual b-f rates but this
remains unimportant for the computational method used in this work.

While collisional rates are important in hotter stellar atmospheres
with high electron densities, they remain nearly negligible when
compared to the radiative rates for the low electron densities found
in cool stars. We have approximated bound-free collisional rates using
the semi-empirical formula of Drawin(1961)\nocite{drawin61}. The
bound-bound collisional rates are approximated by the semi-empirical
formula of Allen (1973)\nocite{allen_aq}, while the Van~Regemorter's
formula (1962\nocite{vr62}) was used for permitted transitions.

A more accurate treatment of this model atom requires the availability
of more accurate collisional and photo-ionization rates for Ti~I. In the 
present calculations we have neglected collisions with particles other than 
electrons because the cross-sections are basically unknown. 
Additional collisional processes would tend to restore LTE, therefore, the 
NLTE effects that we obtain in our calculations should be maximized.

\section{NLTE effects on Ti~I line profiles in M Star Spectra}




 We demonstrate the effects of NLTE on the formation of Ti~I lines in 
Figs.~\ref{spec3} and \ref{spec4}. In general, the NLTE effects are
smaller for the cooler models and they are larger for the giants than 
for the dwarfs at given effective temperature. In the optical spectral region,
  the changes caused by the 
Ti~I NLTE line formation are very small and would be hard to observe due to the
enormous crowding of lines in this spectral region. 

In the cooler models, the Ti~I lines form deeper in the atmosphere in a region 
in which the radiation field is nearly Planckian and thus NLTE effects are 
very small. This is due to the enormous background opacity of TiO and other 
molecules. In the outer atmosphere of the cooler models the concentration of 
Ti~I is much smaller than that of TiO, thus the effect of the large 
departures from LTE that we find in these regions on the Ti~I line 
profiles is very small. 

 In the hotter models, NLTE effects, in particular on the near-IR lines, are 
much larger. In these models the line forming region of the Ti~I lines is 
inside the region in which the departures from LTE are significant. In 
addition, the TiO opacity is relatively smaller than in the cooler models. For 
the  Ti~I line at $\lambda_{\rm vac}\approx9641\ang$, NLTE effects make the core 
of the line deeper than the LTE model predicts. This is due to line 
scattering which removes photons from the line core and re-distributes them 
into the line wings. The same effects are present in most Ti~I lines, but there 
are a few exceptions for which the line is weaker in NLTE than in LTE. 
Abundance determinations of Ti, or likely all metals, from near-IR or IR lines 
should therefore include NLTE effects wherever possible. 

 For the model parameters that we have considered so far, the effects of Ti~I 
NLTE on the TiO bands is very small. This seems to be due to the fact that the 
line forming region of TiO (around $\tstd \approx 10^{-3}$ is dominating the line 
forming region of the Ti~I lines (due to the larger opacity of the TiO 
molecule compared to Ti~I). Therefore, the TiO lines form in a region in which 
the Ti~I atom is basically in LTE and the Ti~I NLTE effects that are important 
at smaller optical depths cannot affect the TiO lines significantly. The 
situation is different for the TiO$^+$ molecule which forms from Ti~II and 
O~I (cf.\ Fig.~\ref{fig3}). 
This molecule is very sensitive to the Ti~I NLTE effects and its 
lines would be helpful indicators for Ti~I NLTE effects. 

 In further investigations it will be very interesting to look for NLTE 
effects in the TiO line formation itself. This is certainly feasible with 
modern numerical techniques once adequate data for TiO are available. 


%

\smallskip
\begin{small}
\noindent{\em Acknowledgements:}
It is a pleasure to thank G. Basri, H. Jones, G. Marcy, S. Starrfield, S. Shore, and
S. Viti for stimulating discussions.
This work was supported in part by NASA LTSA grants to Arizona State
University, by NASA grant NAGW-2999 and NSF grant 
AST-9417242 to University of Oklahoma; as well as NSF and NASA EPSCoR
grants to Wichita State University. Some of the calculations presented
in this paper were performed at the San Diego Supercomputer Center (SDSC),
with support from the National Science Foundation, from the NERSC, and from
the U.S. DoE. We thank all these institutions for a generous allocation of
computer time.
\end{small}
\vskip -12pt

\bibliography{yeti,opacity,novae,ltsa,mdwarf}


\end{document}